\documentclass[11pt]{article}
\usepackage{verbatim}
\makeatletter \@addtoreset{equation}{section} \makeatother

\newcommand{\noi}{\vspace{12pt}\noindent}
\newcommand{\beq}{\begin{equation}}
\newcommand{\eeq}{\end{equation}}
\newcommand{\bea}{\begin{eqnarray}}
\newcommand{\eea}{\end{eqnarray}}

\newcommand{\e}[1]{{(\ref{#1})}}
\newcommand{\eq}[1]{{eq.\ (\ref{#1})}}
\newcommand{\es}[2]{{(\ref{#1}) and (\ref{#2})}}

\newcommand{\Ref}[1]{{Ref.~\cite{#1}}}

\newcommand{\ie}{{${ i.e., \ }$}}
\newcommand{\eg}{{${ e.g., \ }$}}

\newcommand{\cf}{{cf.\ }}

\newcommand{\wrt}{{with respect to }}
\newcommand{\wrtt}{{with respect to the }}

\renewcommand{\~}{ \ }
\renewcommand{\=}{ \ = \ }
\newcommand{\eps}{\varepsilon^{}}

\newcommand{\Tilde}{\widetilde}

\newcommand{\p}{\!{}^{}}
\newcommand{\q}{{}^{}}

\newcommand{\cA}{{\cal A}}

\newcommand{\cO}{{\cal O}}
\newcommand{\cZ}{{\cal Z}}

\newcommand{\sdet}{{\rm sdet}}

\newcommand{\for}{{\rm for}}

\newcommand{\Hf}{{1 \over 2}}

\newcommand{\Ih}{{i \over \hbar}}
\newcommand{\Hi}{{\hbar \over i}}

\newcommand{\twostack}[2]{\begin{array}{c} \lower.8ex\hbox{${#1}$}
                     \cr \raise.8ex\hbox{${#2}$} \end{array}}
\newcommand{\deder}[1]{{ 
 {\stackrel{\raise.2ex\hbox{$\leftarrow$}}{\delta^{r}}   } 
\over {   \delta {#1}}  }}
\newcommand{\dedel}[1]{{ 
 {\stackrel{\lower.3ex \hbox{$\rightarrow$}}{\delta^{\ell}}   }
 \over {   \delta {#1}}  }}
\newcommand{\papar}[1]{{ 
 {\stackrel{\raise.2ex\hbox{$\leftarrow$}}{\partial^{r}}   } 
\over {   \partial {#1}}  }}
\newcommand{\papal}[1]{{ 
 {\stackrel{\lower.3ex \hbox{$\rightarrow$}}{\partial^{\ell}}   }
 \over {   \partial {#1}}  }}
\newcommand{\rpa}[1]{{ 
 \stackrel{\raise.2ex\hbox{$\leftarrow$}}{\partial^{r}_{#1}}   }}
\newcommand{\lpa}[1]{{ 
 \stackrel{\lower.3ex\hbox{$\rightarrow$}}{\partial^{\ell}_{#1}}  }}

\newcommand{\proofbox}{\begin{flushright}
${\,\lower0.9pt\vbox{\hrule \hbox{\vrule
height 0.2 cm \hskip 0.2 cm \vrule height 0.2 cm}\hrule}\,}$
\end{flushright}}

\begin{document}
\thispagestyle{empty}
\title{\Large{\bf Path Integral Formulation with Deformed Antibracket}}
\author{{\sc Igor~A.~Batalin}$^{a}$ and {\sc Klaus~Bering}$^{b}$ \\~\\
$^{a}$I.E.~Tamm Theory Division\\
P.N.~Lebedev Physics Institute\\Russian Academy of Sciences\\
53 Leninsky Prospect\\Moscow 119991\\Russia\\~\\
$^{b}$Institute for Theoretical Physics \& Astrophysics\\
Masaryk University\\Kotl\'a\v{r}sk\'a 2\\CZ--611 37 Brno\\Czech Republic}
\maketitle
\vfill
\begin{abstract}
We propose how to incorporate the Leites-Shchepochkina-Konstein-Tyutin
deformed antibracket into the quantum field-antifield formalism.
\end{abstract}
\vfill
\begin{quote}
PACS number(s): 11.10.-z; 11.10.Ef; 11.15.-q; 11.15.Bt; 11.15.Tk; 11.25.Sq. 
\hrule width 5.cm \vskip 2.mm \noindent 
$^{a}${\small E--mail:~{\tt batalin@lpi.ru}} \hspace{10mm}
$^{b}${\small E--mail:~{\tt bering@physics.muni.cz}} \\
\end{quote}

\newpage

\section{Introduction}
\label{secintro}

\noi
The concept of deformations in the field-antifield formalism
\cite{bv81,bv83,bv84} based on a nilpotent higher-order $\Delta\q_{*}$ operator
was developed in a series of papers
\cite{bt94def,f95,ad96,bda96,bbd96,bbd97,bm98,bm989,bm99}.
Such deformations typically modify the Jacobi identity with BRST-exact terms. 
In contrast, we shall in this paper only discuss {\em local} deformations of
the antibracket with a Grassmann-even deformation parameter such that the
Jacobi identity holds strongly, and without assuming an underlying
$\Delta\q_{*}$ operator a priori. Recently \cite{leites01,kontyu08,kontyu10}, 
a non-trivially deformed antibracket  
\beq
(f,g)\q_{*}\~:=\~(f,g)+(-1)^{\eps_{f}}
(\frac{\kappa c(\kappa)}{1\!+\!\frac{\kappa c(\kappa)}{2}N}\Delta f)
\cdot(1\!-\!\frac{N}{2})g+((1\!-\!\frac{N}{2})f)
\cdot\frac{\kappa c(\kappa)}{1\!+\!\frac{\kappa c(\kappa)}{2}N}\Delta g\~,
\label{ktdeformedantibracket}
\eeq
for functions $f,g$ of finitely many variables $z^{A}$ was constructed inside 
various algebras $\cA$ (\eg polynomial algebra, algebra of smooth functions 
with compact support, etc.). Here $\kappa$ is a deformation parameter;
$c(\kappa)\!=\!\sum_{k=0}^{\infty}c\q_{k}\kappa^{k}$ is an arbitrary formal
power series in $\kappa$; and $N\!:=\!z^{A}\partial/\partial z^{A}$
is the Euler/conformal vector field. Moreover, it was shown \cite{kontyu08}
that this deformed antibracket \e{ktdeformedantibracket} is {\em unique} modulo
trivial deformations and reparametrizations of the deformation parameter
$\kappa$. Thus, it is expected to play a central r\^ole.

\noi
In this paper, we propose how to incorporate the non-trivially\footnote{A
trivial deformation $(f,g)\q_{*}\!=\!T^{-1}(Tf,Tg)$ of the antibracket with
$T\!=\!1+\cO(\kappa)$ amounts to a trivial deformation 
$\Delta\q_{*}\!=\!T^{-1}\Delta T$ and $f*g\!=\!T^{-1}(Tf\cdot Tg)$
of the underlying BV algebra $(\cA;\Delta\q_{*};*)$.}
deformed antibracket \e{ktdeformedantibracket} into the quantum field-antifield
formalism \cite{bv81,bv83,bv84}. Concretely, we suggest a $\kappa$-deformed 
odd Laplacian; quantum master action $W\!=\!S\!+\!\cO(\hbar)$; quantum master
equation; and partition function $\Tilde{\cZ}$, such that, the classical master
equation is given in terms of the above $\kappa$-deformed antibracket
\beq
(S,S)\q_{*}\=0 \~; \label{cme01}
\eeq
the classical BRST symmetry is $s\!=\!(S,\cdot)\q_{*}$; and the partition
function $\Tilde{\cZ}$ is formally independent of the gauge-fixing $X$.

\noi
How would a $\kappa$-deformation be realized in practice? {}Firstly, we stress 
that field theory implies infinitely many $z^{A}$-variables, so that both the
Euler vector field $N$ and the odd Laplacian $\Delta$ would need
regularization. Nevertheless, it is reasonable to assume that the naive 
finite-dimensional
$N$-deformation \e{ktdeformedantibracket} still serves as a model of what to
come in field theory. Secondly, we note that the traditional field-antifield
approach \cite{bv81,bv83,bv84} (where one starts from a classical action, which
is independent of ghosts and antifields, and one introduces ghosts and 
antifields as generators of gauge- and BRST-symmetry, respectively) is 
{\em not} expected to produce a $\kappa$-deformation, as the antibracket
traditionally remains on Darboux form. Rather, a relevant physical system
should have an antisymplectic phase space built in from the beginning, like,
\eg closed string field theory \cite{zwiebach93}, or generalized Poisson sigma
models \cite{aksz97,bm01,bm02}. It is believed that the $\kappa$-deformation
here could be caused by a choice of regularization scheme that manifestly
preserves the Jacobi identity.

\noi
The new construction is motivated by two key ideas, which may be symbolized
with the introduction of a Bosonic and Fermionic variable, $t$ and $\theta$,
respectively, with collective notation $\tau\!:=\!\{t;\theta\}$. 
Mathematically, they are, in fact, intimately tied to Lie
cohomology theory. We will only here sketch the Lie cohomology argument, and
defer a more detailed explanation to an accompanying paper \cite{bb10}. Recall
that the ambiguity/uniqueness of deformations of a Lie-bracket is measured by
the second Lie cohomology group, while the first Lie cohomology group
classifies outer(=non-Hamiltonian) Lie algebra derivations. Konstein and Tyutin
have calculated \cite{kontyu08} the first and second Lie cohomology group for
the constant, non-degenerated antibracket $(\cdot,\cdot)$. The first Lie
cohomology group is two-dimensional, and, in detail, it is generated by the odd
Laplacian $\Delta$ and the affine operator $N\!-\!2$. The second Lie cohomology
group is two-dimensional as well, and, in accordance with the K\"unneth 
formula, it is generated by all possible non-zero\footnote{The last 
$\Delta\cup\Delta\!=\!0$ of the $2\!\times\!2\!=\!4$ possibilities vanishes 
identically, because the cup product $\cup$ is (graded) commutative.} cup
product combinations of the first cohomology. These are
$\Delta\cup(N\!-\!2)\!=\!(N\!-\!2)\cup\Delta$ and $(N\!-\!2)\cup(N\!-\!2)$,
which lead to two deformed antibrackets, with an even and an odd deformation
parameter, respectively, where we here will only consider the former. The first
key idea is to {\em suspend} the algebra $\cA$ by introducing a suspension
parameter $t$ to turn the affine operator $N\!-\!2$ into a genuine vector field
$N\p_{\tau}\!=\!N\!+\!t\partial/\partial t$, which satisfies the Leibniz rule.
The non-triviality of the  $N\p_{\tau}$ vector field in the $\{z^{A};t\}$ space
means that it is {\em not} a Hamiltonian vector field. The second key idea is
to complement the $\{z^{A};t\}$ space with an antisymplectic partner $\theta$,
in such a way, that $\theta$ becomes (minus) the Hamiltonian generator for the
vector field $N\p_{\tau}\!=\!-(\theta,\cdot)\q_{\tau}$, and hence, so that the 
vector field $N\p_{\tau}$ becomes trivial, and, in turn, it makes the 
corresponding $(t;\theta)$-extended deformed antibracket
$(\cdot,\cdot)\q_{\tau*}$ trivial.

\section{Basic Setting: Constant Non-Degenerate Antibracket}
\label{secnondegoddpoisalg}

\noi
Let $\cA\!:=\!C[[z]]$ be the algebra of formal power series $f\!=\!f(z)$ in
$2n$ variables $z^{A}$ of Grassmann parity $\eps(z^{A})\!\equiv\!\eps_{A}$,
equipped with a constant, non-degenerate antibracket $E^{AB}\!=\!(z^{A},z^{B})$
with Grassmann parity $\eps(E^{AB})\!=\!\eps_{A}\!+\!1\!+\!\eps_{B}$ 
corresponding to the odd Laplacian
\beq
\Delta\~:=\~\frac{(-1)^{\eps_{A}}}{2}\papal{z^{A}}E^{AB}\papal{z^{B}}\~,\qquad 
\Delta^{2}\= 0\~,\qquad \eps(\Delta)\=1\~.
\eeq
The antibracket
\beq
(f,g)\~:=\~(-1)^{\eps_{f}}[[\stackrel{\rightarrow}{\Delta},f],g]1 
\=-(-1)^{(\eps_{f}+1)(\eps_{g}+1)}(g,f) 
\~,\qquad f,g\~\in\~\cA\~, \label{wittenformula}
\eeq
satisfies skewsymmetry \e{wittenformula}, the Jacobi identity
\beq
\sum_{{\rm cycl.}\~f,g,h}(-1)^{(\eps_{f}+1)(\eps_{h}+1)}(f,(g,h))\=0\~,
\qquad f,g,h\~\in\~\cA\~, \label{jacid}
\eeq
and the Leibniz rule/Poisson property
\beq
(fg,h)\=f(g,h)+(-1)^{\eps_{f}\eps_{g}}g(f,h)\~,\qquad f,g,h\~\in\~\cA\~.
\label{leibniz}
\eeq

\section{Non-Trivially Deformed Algebra $\cA$}
\label{secnontrivdeformoddpoisalg}

\noi
We will from now on use the simplifying convention that the power series from
\eq{ktdeformedantibracket} is $c(\kappa)\!=\!-2$. To reintroduce the whole 
$c(\kappa)$ series, just replace $\kappa\to -\frac{\kappa c(\kappa)}{2}$. The 
deformed odd Laplacian $\Delta\q_{*}$ and antibracket $(\cdot,\cdot)\q_{*}$,
\cf \eq{ktdeformedantibracket}, read
\bea
\Delta\q_{*}&:=&\Delta\frac{1}{1\!-\!K}\=\frac{1}{1\!-\!\kappa N}\Delta
\~,\qquad \Delta^{2}_{*}\=0\~,
\label{deltastar} \\
(f,g)\q_{*}&:=&(f,g)+(-1)^{\eps_{f}}(\Delta\q_{*}f)\cdot(K g)
+(Kf)\cdot(\Delta\q_{*}g) \label{ktdeformedantibracket02} \\
&=&(-1)^{\eps_{f}}\Delta(fg)
-(1\!-\!K)\{(-1)^{\eps_{f}}(\Delta\q_{*}f)g+f(\Delta\q_{*}g)\}
\label{wagelywittenlike} \\
&=&-(-1)^{(\eps_{f}+1)(\eps_{g}+1)}(g,f)\q_{*}\~,\qquad f,g\~\in\~\cA\~,  \\
K&:=&\kappa(N\!-\!2)\~, \qquad N\~:=\~z^{A}\papal{z^{A}}\~,\qquad 
[\Delta,N]\=2\Delta\~,\qquad\eps(\kappa)\=0\~. \label{kay}
\eea 
Within the algebra $\cA$, the deformed odd Laplacian $\Delta\q_{*}$ is 
characterized by nilpotency, and the property 
\beq
\Delta\q_{*}(f,g)\q_{*}\=(\Delta\q_{*}f,g)\q_{*}
-(-1)^{\eps_{f}}(f,\Delta\q_{*}g)\q_{*}\~,\qquad f,g\~\in\~\cA\~,
\eeq
\ie that $\Delta\q_{*}$ differentiates the deformed antibracket
$(\cdot,\cdot)\q_{*}$. The standard Witten formula \e{wittenformula}, 
\cf \Ref{witten90}, is deformed into \e{wagelywittenlike}, which, in turn,
can be used to prove the Jacobi identity \e{deformjacid} for the deformed
antibracket $(\cdot,\cdot)\q_{*}$,
\beq
\sum_{{\rm cycl.}\~f,g,h}(-1)^{(\eps_{f}+1)(\eps_{h}+1)}(f,(g,h)\q_{*})\q_{*}
\=0\~,\qquad f,g,h\~\in\~\cA\~. \label{deformjacid}
\eeq
Note that the deformed antibracket $(\cdot,\cdot)\q_{*}$ does {\em not} satisfy
the Leibniz rule/Poisson property, \cf \eq{leibniz}, and hence the deformed 
antibracket $(\cdot,\cdot)\q_{*}$ is, technically speaking, {\em not} an odd 
Poisson bracket. Therefore, the deformation and the corresponding cohomology
must be treated within the framework of (infinite-dimensional, graded) 
Lie algebras instead of (finitely generated, graded) Poisson algebras.

\section{$k$-Suspended Deformed Operators}
\label{secgendefop}

\noi
Define for later convenience a $k$-suspended deformed odd Laplacian
$\Delta^{(k)}_{*}$ and a $(k,\ell)$-suspended deformed antibracket
$(\cdot,\cdot)^{(k,\ell)}_{*}$,
\bea
\Delta^{(k)}_{*}&:=&\Delta\frac{1}{1\!-\!K^{(k)}}\~,\qquad 
(\Delta^{(k)}_{*})^{2}\=0\~,\qquad K^{(k)}\Delta\=\Delta K^{(k-2)}\~,
\label{kdeltastar} \\
(f,g)^{(k,\ell)}_{*}
&:=&(f,g)+(-1)^{\eps_{f}}(\Delta^{(k)}_{*}f)\cdot(K^{(\ell)}g)
+(K^{(k)}f)\cdot(\Delta^{(\ell)}_{*}g) \\
&=&(-1)^{\eps_{f}}\Delta(fg)
-(1\!-\!K^{(k+\ell+2)})\{(-1)^{\eps_{f}}(\Delta^{(k)}_{*}f)g
+f(\Delta^{(\ell)}_{*}g)\} \label{wagelywittenlike01} \\
&=&-(-1)^{(\eps_{f}+1)(\eps_{g}+1)}(g,f)^{(\ell,k)}_{*}\~,
\qquad f,g\~\in\~\cA\~, \\
K^{(k)}&:=&\kappa N^{(k)}\~,\qquad N^{(k)}\~:=\~N\!+\!k\~,\qquad 
N^{(k)}_{*}\~:=\~N^{(k)}\frac{1}{1\!-\!K^{(k)}}\~,\qquad 
K^{(k)}_{*}\~:=\~\kappa N^{(k)}_{*}\~,\qquad  \label{kaykay}
\eea
where $k,\ell$ are integers. In particular, the $k$-suspended definitions
\e{kdeltastar}-\e{kaykay} generalize the definitions \e{deltastar}-\e{kay} of
Section~\ref{secnontrivdeformoddpoisalg} in the following way,
\beq
\Delta^{(-2)}_{*}\~\equiv\~\Delta\q_{*}\~,\qquad 
(f,g)^{(-2,-2)}_{*}\~\equiv\~(f,g)\q_{*}
\~,\qquad K^{(-2)}\~\equiv\~K\~,\qquad N^{(0)}\~\equiv\~N\~.
\eeq
Equation \e{wagelywittenlike01} is a $(k,\ell)$-suspended deformed Witten
formula \cite{witten90}. Note also the elementary, but useful, formula 
\beq
K^{(k+\ell)}(fg)\=(K^{(k)}f)g+f(K^{(\ell)}g)\~,\qquad f\~\in\~\cA\~.
\label{useful}
\eeq
Equations \es{wagelywittenlike01}{useful} can be used to prove the Jacobi 
identity
\beq
\sum_{{\rm cycl.}\~(f,k),(g,\ell),(h,m)}(-1)^{(\eps_{f}+1)(\eps_{h}+1)}
(f,(g,h)^{(\ell,m)}_{*})^{(k,\ell+m+2)}_{*}\=0\~,
\qquad f,g,h\~\in\~\cA\~, \label{deformkljacid}
\eeq
and the differentiation rule
\beq
\Delta^{(k+\ell+2)}_{*}(f,g)^{(k,\ell)}_{*}
\=(\Delta^{(k)}_{*}f,g)^{(k+m,\ell)}_{*}
-(-1)^{\eps_{f}}(f,\Delta^{(\ell)}_{*}g)^{(k,m+\ell)}_{*}\~,
\qquad f,g\~\in\~\cA\~.
\eeq

\section{$\tau$-Extended Algebra $\cA\q_{\tau}$}
\label{sectauoddpoisalg}

\noi
Let us now introduce a $\tau$-extended algebra 
$\cA\q_{\tau}\!:=\!C[[z;t;\theta]][\frac{1}{t}]$ of formal (lower truncated) 
Laurent series 
\beq
F\=\sum_{k=-M_{F}}^{\infty}F\p_{(k)}(z;\theta)t^{k}\~,\qquad
F\p_{(k)}(z;\theta)\=F\p_{(k|0)}(z)+\theta F\p_{(k|1)}(z)\~,
\eeq
where the lower limit $k\!=\!-M_{F}$ may depend on the series $F$, and 
$\tau\!:=\!\{t;\theta\}$ is a collective notation for the two new variables
$t$ and $\theta$ of Grassmann parity $\eps(t)\!=\!0$ and $\eps(\theta)\!=\!1$,
respectively. One introduces a suspension map
$\lfloor\cdot\rfloor:\cA\to\cA\q_{\tau}$ as
\beq
\lfloor f\rfloor\~:=\~\frac{f}{t^{2}}\~,\qquad f\~\in\~\cA\~. 
\eeq
The residue map $\pi:\cA\q_{\tau}\!\to\!\cA$ reads
$\pi(F):=\oint_{0}\!\frac{tdt}{2\pi i}\int\!d\theta\~\theta\~F=F\p_{(-2|0)}$
with Berezin integral convention $\int\! d\theta\~\theta\!=\!1$. One has
$\pi\circ\lfloor\cdot\rfloor\!=\!{\rm id}_{\cA}$, or equivalently,
$\pi\circ\lfloor f\rfloor\!=\!f$ for $f\!\in\!\cA$.

\section{$\tau$-Extended Antisymplectic Structure}
\label{sectauoddpoisstructure}

\noi
Define generalized Darboux\footnote{{\em Generalized Darboux coordinates} are
coordinates in which the (odd) Poisson bi-vector is constant, \cf 
\eq{darbouxxx}.} coordinates $\{z^{A}_{0};t\q_{0};t^{*}_{0}\}$ as
\beq
z^{A}_{0}\~:=\~\frac{z^{A}}{t}\~,\qquad t\q_{0}\~:=\~\ln(t)\~,\qquad
t^{*}_{0}\~:=\~\theta\~, 
\eeq 
with inverse transformation
\beq
z^{A}\= e^{t\q_{0}} z^{A}_{0}\~,\qquad t\=e^{t\q_{0}}\~,\qquad 
\theta\=t^{*}_{0}\~.
\eeq
The Berezin volume densities for the generalized Darboux and original 
coordinates are chosen as
\beq
\rho\q_{0}\~:=\~1\~,\qquad
\rho\q_{\tau}\~:=\~\frac{\rho\q_{0}}{J}\=\frac{1}{t}\~, \qquad 
J\~:=\~\sdet\frac{\partial\{z^{A};t;\theta\}}
{\partial\{z^{A}_{0};t\q_{0};t^{*}_{0}\}}\=t \~.\label{berezinvolume}
\eeq
The algebra $\cA\q_{\tau}$ is equipped with the second-order odd 
Laplacian\footnote{Theoretically, the parameter $t$ serves as a {\em unit} of
suspension. In practice, it may be more convenient to expand in terms of
its square $t\q_{2}\!:=\!t^{2}$, so that $\lfloor f\rfloor\!:=\!f/t\q_{2}$; 
$N\p_{\tau}\!:=\!N\!+\!2t\q_{2}\partial/\partial t\q_{2}$; 
$\Delta\q_{\tau}\!:=\!t\q_{2}\Delta\!+\!N\p_{\tau}\partial/\partial\theta$; 
etc.}
\bea
\Delta\q_{\tau}&:=&
\frac{(-1)^{\eps_{A}}}{2}\papal{z^{A}_{0}}E^{AB}\papal{z^{B}_{0}}
+\papal{t\q_{0}}\papal{t^{*}_{0}}
\=t^{2}\Delta+N\p_{\tau}\papal{\theta}\~,\qquad 
\Delta^{2}_{\tau}\= 0\~, \\
N\p_{\tau}&:=&N+t\papal{t}\=-(\theta,\cdot)\q_{\tau}\~,\qquad 
[N\p_{\tau},\Delta\q_{\tau}]\=0\~, \\
(F,G)\q_{\tau}
&:=&(-1)^{\eps_{F}}[[\stackrel{\rightarrow}{\Delta}\p_{\tau},F],G]1\~,
\eea
such that the suspension map $\lfloor\cdot\rfloor$ intertwines between an
operation and its $\tau$-extended counterpart,
\bea
\Delta\q_{\tau}\lfloor f\rfloor&=&\Delta f\~,\qquad 
N\p_{\tau}\lfloor f\rfloor\=\lfloor(N\!-\!2)f\rfloor
\~,\qquad f\~\in\~\cA\~, \\
(\lfloor f\rfloor,\lfloor g\rfloor)\q_{\tau}
&=&\lfloor (f,g)\rfloor\~,\qquad
(f,\lfloor g\rfloor)\q_{\tau}\= (f,g)\~,\qquad f,g\~\in\~\cA\~.
\eea
The non-vanishing antibrackets $(\cdot,\cdot)\q_{\tau}$ of the fundamental
variables $\{z^{A};t;\theta\}$
read
\beq
(z^{A},z^{B})\q_{\tau}\=t^{2}E^{AB}\~,\qquad (z^{A},\theta)\q_{\tau}\=z^{A}\~,
\qquad (t,\theta)\q_{\tau}\=t\~,
\eeq
or in terms of generalized Darboux coordinates
$\{z^{A}_{0};t\q_{0};t^{*}_{0}\}$,
\beq
(z^{A}_{0},z^{B}_{0})\q_{\tau}\=E^{AB}\~,\qquad 
(t\q_{0},t^{*}_{0})\q_{\tau}\=1\~.
\label{darbouxxx}
\eeq

\section{Trivially Deformed $\tau$-Extended Odd Poisson Algebra $\cA\q_{\tau}$}
\label{sectrivdeformtauoddpoisalg}

\noi
Define a trivially deformed odd Laplacian
\bea
\Delta\q_{\tau*}&:=&\Delta\q_{\tau}\frac{1}{1\!-\!K\q_{\tau}}
\=T^{-1}\Delta\q_{\tau}T\~, \qquad\Delta^{2}_{\tau*}\=0\~, 
\label{deformedtaudelta} \\
K\q_{\tau}&:=&\kappa N\q_{\tau}\~,\qquad [K\q_{\tau},\Delta\q_{\tau}]\=0\~, 
\eea
\cf Appendix~\ref{appdeformedtaudelta}, where $T$ is the trivialization map
in the $\tau$-extended algebra $\cA\q_{\tau}$,
\beq
T\~:=\~1+\kappa\theta\Delta\q_{\tau*}\~,\qquad
T^{-1}\~:=\~1-\kappa\theta\Delta\q_{\tau}\~,\qquad T^{-1}T\=1\=TT^{-1}\~, 
\label{ttrelation}
\eeq
\cf Appendix~\ref{appttrelation}, so that in the suspended sector,
\beq
\Delta\q_{\tau*}\lfloor f\rfloor\=\Delta\q_{*}f\~,\qquad 
K\q_{\tau}\lfloor f\rfloor\=\lfloor Kf\rfloor\~,\qquad f\~\in\~\cA\~.
\eeq
If one expands \wrtt $t$ variable, one gets
\bea
\Delta\q_{\tau*}F&=&\sum_{k}\left(t^{2}\Delta^{(k)}_{*}F\p_{(k)}
+N^{(k)}_{*}F\p_{(k|1)}\right)t^{k}
\~,\qquad F\~\in\~\cA\q_{\tau}\~, \\
K\q_{\tau}F&=&\sum_{k}(K^{(k)}F\p_{(k)})t^{k}\~,\qquad F\~\in\~\cA\q_{\tau}\~.
\eea
Define a trivially deformed antibracket
\bea
(F,G)\q_{\tau*}&:=&T^{-1}(TF,TG)\q_{\tau} 
\=(F,G)\q_{\tau}+(-1)^{\eps_{F}}(\Delta\q_{\tau*}F)\cdot K\q_{\tau}G
+(K\q_{\tau}F)\cdot\Delta\q_{\tau*}G  \label{deformtauantibracket} \\
&=&(-1)^{\eps_{F}}\Delta\q_{\tau}(FG)
-(1\!-\!K\q_{\tau})\{(-1)^{\eps_{F}}(\Delta\q_{\tau*}F)G
+F\Delta\q_{\tau*}G\} \\
&=&-(-1)^{(\eps_{F}+1)(\eps_{G}+1)}(G,F)\q_{\tau*}
\~,\qquad F,G\~\in\~\cA\q_{\tau}\~,\label{wagelywittenliketau}
\eea
\cf Appendix~\ref{appdeformtauantibracket}, so that in the suspended sector,
\beq
(\lfloor f\rfloor,\lfloor g\rfloor)\q_{\tau*}
\=\lfloor (f,g)\q_{*}\rfloor\~,\qquad f,g\~\in\~\cA\~.\label{biggestsuccess}
\eeq
If one expands \wrtt $t$ variable, one gets
\bea
(F,G)\q_{\tau*}&=&\sum_{k,\ell}\left(
t^{2}(F\p_{(k)},G\p_{(\ell)})^{(k,\ell)}_{*}
+(-1)^{\eps_{F}} (\frac{1}{1\!-\!K^{(k)}}F\p_{(k|1)})
\cdot N^{(\ell)}G\p_{(\ell)}\right. \cr
&&\left. +(N^{(k)}F\p_{(k)})\cdot
\frac{1}{1\!-\!K^{(\ell)}}G\p_{(\ell|1)}\right)t^{k+\ell}
\~,\qquad F,G\~\in\~\cA\q_{\tau}\~.
\eea
The trivially deformed antibracket $(\cdot,\cdot)\q_{\tau*}$ satisfies the 
Jacobi identity,
\beq
\sum_{{\rm cycl.}\~F,G,H}(-1)^{(\eps_{F}+1)(\eps_{H}+1)}
(F,(G,H)\q_{\tau*})\q_{\tau*}
\=0\~,\qquad F,G,H\~\in\~\cA\q_{\tau}\~. \label{deformtaujacid}
\eeq
Equation \e{biggestsuccess} therefore gives an alternative derivation of the 
Jacobi identity \e{deformjacid}. Define a trivial associative and commutative 
{\em star product} as
\beq
F*G\~:=\~T^{-1}(TF\cdot TG)
\=FG-(-1)^{\eps_{F}}\kappa\theta(F,G)\q_{\tau*}\~,\qquad F,G\~\in\~\cA\q_{\tau}
\~,\qquad \eps(*)\=0\~, \qquad  \label{trivstarproduct} 
\eeq
\cf Appendix~\ref{apptrivstarproduct}, so that in the suspended sector,
\beq
\lfloor f\rfloor*\lfloor g\rfloor\=\lfloor\lfloor fg\rfloor\rfloor
-(-1)^{\eps_{f}}\kappa\theta\lfloor (f,g)\q_{*}\rfloor\~,
\qquad f,g\~\in\~\cA\~.
\eeq
The trivially deformed Witten formula \cite{witten90} reads
\beq
(F,G)\q_{\tau*}\=(-1)^{\eps_{F}}\Delta\q_{\tau*}(F*G)
-(-1)^{\eps_{F}}(\Delta\q_{\tau*}F)*G-F*\Delta\q_{\tau*}G
\~,\qquad F,G\~\in\~\cA\q_{\tau}\~.
\eeq
The Leibniz rule/Poisson property reads
\beq
(F*G,H)\q_{\tau*}\=F*(G,H)\q_{\tau*}+(-1)^{\eps_{F}\eps_{G}}G*(F,H)\q_{\tau*}
\~,\qquad F,G,H\~\in\~\cA\q_{\tau}\~.
\eeq
The Getzler identity \cite{g94} for the BV algebra 
$(\cA\q_{\tau};\Delta\q_{\tau*};*)$ reads
\bea
0&=&\Delta\q_{\tau*}(F*G*H)-\Delta\q_{\tau*}(F*G)*H
-(-1)^{\eps_{F}}F*\Delta\q_{\tau*}(G*H)
-(-1)^{\eps_{G}\eps_{H}}\Delta\q_{\tau*}(F*H)*G \cr
&&+(\Delta\q_{\tau*}F)*G*H+(-1)^{\eps_{F}}F*(\Delta\q_{\tau*}G)*H
+(-1)^{\eps_{F}+\eps_{G}}F*G*\Delta\q_{\tau*}H
\~,\qquad F,G,H\~\in\~\cA\q_{\tau}\~. \cr &&
\eea
which encodes the vanishing of higher antibrackets \cite{bda96,bbd96,b06}. 
The {\em star exponential} is defined as
\bea
e^{B}_{*}&:=&1+B+\Hf B*B+\frac{1}{3!}B*B*B+\frac{1}{4!}B*B*B*B+\ldots 
\=T^{-1}e^{(TB)} \cr
&=&e^{B}\left(1-\Hf\kappa\theta(B,B)\q_{\tau*}\right)
\=e^{B-\Hf\kappa\theta(B,B)\q_{\tau*}}
\~,\qquad B\~\in\~\cA\q_{\tau}\~,\qquad \eps(B)\=0\~,\label{starexpdef}
\eea
\cf Appendix~\ref{appstarexpdef}. The star exponential satisfies
\bea
e^{-B}_{*} * e^{B}_{*}&=&1\~,\qquad
e^{-B}_{*}*(\Delta\q_{\tau*}e^{B}_{*})
\=(\Delta\q_{\tau*}B)+\Hf (B,B)\q_{\tau*}\~,  \qquad 
\delta e^{B}_{*}\=e^{B}_{*}*\delta B\~,  \\
e^{B+B^{\prime}}_{*}&=&e^{B}_{*}*e^{B^{\prime}}_{*}\~,\qquad 
B,B^{\prime}\~\in\~\cA\q_{\tau}\~,\qquad \eps(B)\=0\=\eps(B^{\prime})\~.
\eea
If we want to stress the deformation parameter $\kappa$, we write a subindex 
``$(\kappa)$'', \ie
\beq
T\~\equiv\~T\p_{(\kappa)}\~,\quad
\Delta\q_{\tau*}\~\equiv\~\Delta\q_{\tau*(\kappa)}\~,\quad
(\cdot,\cdot)\q_{\tau*}\~\equiv\~(\cdot,\cdot)\q_{\tau*(\kappa)}\~,\quad
F*G\~\equiv\~F*\p_{(\kappa)}G\~,\quad
e^{B}_{*}\~\equiv\~e^{B}_{*(\kappa)}\~.\quad
\eeq

\section{Deformed Quantum Master Equations}
\label{secdeformedqme02}

\noi
We will here for simplicity use the strong first-level\footnote{
The strong first-level gauge-fixing action $\Tilde{X}$ also depends on 
first-level Lagrange multipliers
$\{\lambda^{\Tilde{\alpha}}\}\!=\!\{\lambda^{\alpha};\lambda\q_{\theta}\}$, and
is capable of incorporating all Abelian gauge-fixing constraints 
$(G\q_{\Tilde{\alpha}},G\q_{\Tilde{\beta}})\q_{\tau}=0$. {}For non-Abelian 
gauge-fixing constraints, it is necessary to add weak terms in the quantum 
master equation \cite{bms95}, or still better, to go to the second-level 
formalism, which introduces antifields $\lambda^{*}_{\Tilde{\alpha}}$ for the 
first-level Lagrange multipliers; second-level Lagrange multipliers 
$\lambda_{(2)}^{\Tilde{\alpha}}$; odd Laplacian 
$\Delta\q_{[1]\tau*}=\Delta\q_{\tau*}
+(-1)^{\eps_{\tilde{\alpha}}}\partial/\partial\lambda^{\Tilde{\alpha}}\~
\partial/\partial\lambda^{*}_{\Tilde{\alpha}}$; and action
$\Tilde{W}\q_{[2]}=\lambda^{*}_{\Tilde{\alpha}}\lambda_{(2)}^{\Tilde{\alpha}}
+\Tilde{W}$.} $W$-$X$-formalism, which consists of a gauge-generating and a
gauge-fixing action, $W$ and $X$
\cite{bt92,bt93,bt94,bms95,bt96,bbd96,bbd06,bb07}. In the $\tau$-extended case,
we adorn the two actions with tildes. The two quantum master equations are
\beq
\Delta\q_{\tau*(\kappa)}e^{\Ih \Tilde{W}}_{*(\kappa)}\=0\~,\qquad
\Delta\q_{\tau*(-\kappa)}e^{\Ih \Tilde{X}}_{*(-\kappa)}\=0\~,\qquad 
\Tilde{W}, \Tilde{X}\~\in\~\cA\q_{\tau}\~,\qquad 
\eps(\Tilde{W})\=0\=\eps(\Tilde{X})\~, \label{tildeexpqme}
\eeq
or equivalently,
\beq
\Hf(\Tilde{W},\Tilde{W})\q_{\tau*(\kappa)}
\=i\hbar\Delta\q_{\tau*(\kappa)}\Tilde{W}\~,\qquad
\Hf(\Tilde{X},\Tilde{X})\q_{\tau*(-\kappa)}
\=i\hbar\Delta\q_{\tau*(-\kappa)}\Tilde{X}\~.\label{tildeaddqme}
\eeq
{}From now on, it is implicitly assumed that the star deformations in the
$\Tilde{W}$- and $\Tilde{X}$-sector refer to the deformation parameter $\kappa$
and $-\kappa$, respectively, to avoid clutter. Consider first the $\Tilde{W}$
action. Let us mention that $\Tilde{W}$ satisfies the $\kappa$-deformed quantum
master equation if and only if $T\Tilde{W}$ satisfies the undeformed quantum
master equation. If one expands the quantum master equation for
$\Tilde{W}\!=\!\sum_{k=-\infty}^{\infty}\Tilde{W}\p_{(k)}t^{k}$ \wrtt $t$
variable, one gets
\bea
\Hf\sum_{\ell=-\infty}^{\infty}
(\Tilde{W}\p_{(\ell)},\Tilde{W}\p_{(k-\ell)})^{(\ell,k-\ell)}_{*}
&+&\sum_{\ell=-\infty}^{\infty}N^{(\ell)}\Tilde{W}\p_{(\ell)}
\cdot\frac{1}{1\!-\!K^{(k-\ell+2)}}\Tilde{W}\p_{(k-\ell+2|1)}  \cr 
&=&i\hbar\Delta^{(k)}_{*}\Tilde{W}\p_{(k)}
+i\hbar N^{(k+2)}_{*}\Tilde{W}\p_{(k+2|1)}\~.\label{qmet01}
\eea
We next identify the component $\Tilde{W}\p_{(-2|0)}\!=\!S$ with the 
proper\footnote{An action is called {\em proper} (\wrt a set of 
antisymplectic variables) if its corresponding Hessian has rank equal to
half the number of variables at the stationary surface, see \eg \Ref{bb09}.} 
classical action $S$ from \eq{cme01}. To have the classical master equation
\e{cme01} within the $t$-hierarchy \e{qmet01}, the Laurent series 
$\Tilde{W}$ must truncate from below as
\beq
\Tilde{W}\=\sum_{k=-2}^{\infty}\Tilde{W}\p_{(k|0)}t^{k}
+\theta\sum_{k=1}^{\infty}\Tilde{W}\p_{(k|1)}t^{k}\~.
\eeq
The minimal Ansatz for the gauge-generating and gauge-fixing actions,
$\Tilde{W}$ and $\Tilde{X}$ reads\footnote{Note that while the leading term
$\lfloor S\rfloor$ in the $\Tilde{W}$ action is proper in the original
antisymplectic phase space $\{z^{A}\}$, it is in general {\em not} proper in
the $\tau$-extended antisymplectic phase space $\{z^{A};t;\theta\}$. Thus if
one would like to treat the $t$ variable perturbatively, it is necessary to
include $t$-dependent classical (=$\hbar$-independent) terms in the $\Tilde{W}$
action, which necessarily must violate the minimal Ansatz \e{bcw}. We analyze
here the minimal Ansatz \e{bcw} for simplicity, as the Ansatz is consistent
with the quantum master equation \e{tildeaddqme}, but with the caveat that $t$
may acquire a non-perturbative status. }
\bea
\Tilde{W}&=&\frac{1}{t^{2}}W(z;\hbar t^{2};\kappa)
\=\lfloor S \rfloor+\hbar M\q_{1}+\cO(\hbar^{2}t^{2})\~, \qquad 
\frac{\partial\Tilde{W}}{\partial\theta}=0\~,\label{bcw}\\
\Tilde{X}&=&X(\frac{z}{t};\lambda;\hbar)+i\hbar\theta\lambda\q_{\theta}
\=X(z\q_{0};\lambda;\hbar)+i\hbar t^{*}_{0} \lambda\q_{0}\~,\qquad 
N\p_{\tau}\Tilde{X}\=0\~, \label{bcx}
\eea
where $\lambda\q_{\theta}\!\equiv\!\lambda\q_{0}$ is a Fermionic 
first-level Lagrange multiplier to gauge-fix the $\theta$ variable,
and where 
\beq
W\=W(z;\hbar t^{2};\kappa)
\=S+\sum_{k=1}^{\infty}(t^{2}\hbar)^{k}M\q_{k}\~,\label{wsm01} \qquad
S\=S(z;\kappa)\~,\qquad M\q_{k}\=M\q_{k}(z;\kappa)\~\~\for\~\~ k\geq 1\~. 
\eeq
In $t$-components, the minimal Ansatz \e{bcw} for $\Tilde{W}$ reads
\beq
\Tilde{W}\p_{(-2)}\=S\~, \qquad 
\Tilde{W}\p_{(2k-2)}\=\hbar^{k}M\q_{k}\~\~\for\~\~ k\geq 1\~, \qquad 
\Tilde{W}\p_{(-2k)}\=0\~\~\for\~\~ k\geq 2\~, \qquad 
\Tilde{W}\p_{(2k+1)}\=0\~.
\eeq
The quantum hierarchy \e{qmet01} for $\Tilde{W}$ becomes
\bea
(S,S)\q_{*}&=&0\~, \label{cme02} \qquad\qquad
(M\q_{1},S)^{(0,-2)}_{*}\=i\Delta\q_{*}S\~, \label{emome} \\
(M\q_{k},S)^{(2k-2,-2)}_{*}&=&i\Delta^{(2k-4)}_{*}M\q_{k-1}
-\Hf\sum_{\ell=1}^{k-1}(M\q_{\ell},M\q_{k-\ell})^{(2\ell-2,2k-2\ell-2)}_{*}
\~\~\for\~\~k\geq 2\~. \label{genmk}
\eea
The hierarchy \e{cme02}-\e{genmk} successively determines $S$ and $M\q_{k}$ for
$k\!\geq\!1$. The untilded gauge-fixing action $X$ satisfies an ordinary
quantum master equation
\beq
\Delta e^{\Ih X}\=0 \qquad  \Leftrightarrow \qquad \Hf(X,X)\=i\hbar\Delta X\~,
\eeq
which is undeformed in the deformation parameter $-\kappa$.

\section{Deformed Path Integral}
\label{secpathintdeform02}

\noi
The first-level {\em path integral measure} is
\beq
d\mu\=\rho\q_{\tau}\~dt\~d\theta\~d\lambda\q_{\theta}[dz][d\lambda]
\=\rho\q_{0}\~dt\q_{0}\~dt^{*}_{0}\~d\lambda\q_{0}[dz_{0}][d\lambda]\~,
\eeq
\cf \eq{berezinvolume}. The {\em transposed operator} $A^{T}$ of an operator 
$A$ is defined via \cite{bbd96}
\beq
\int\!d\mu\~(A^{T}F)\cdot G
\=(-1)^{\eps_{A}\eps_{F}}\int\! d\mu\~F\cdot (AG)\~,
\eeq
where $F,G$ are two arbitrary functions. The transposed odd Laplacians and
transposed Euler vector fields are
\beq
\Delta^{T}\=\Delta\~,\qquad N^{T}\=-N\~,\qquad
\Delta^{T}_{\tau}\=\Delta\q_{\tau}\~,\qquad N^{T}_{\tau}\=-N\p_{\tau}\~,\qquad
\Delta^{T}_{\tau*(\kappa)}\=\Delta\q_{\tau*(-\kappa)}\~.\label{vartransposedop}
\eeq
The first-level {\em path integral} $\Tilde{\cZ}$ in the $\tau$-extended
antisymplectic phase space is defined as
\beq
\Tilde{\cZ}
\=\int d\mu\~e^{\Ih\Tilde{W}}_{*(\kappa)}\cdot e^{\Ih\Tilde{X}}_{*(-\kappa)} 
\=\int d\mu\~e^{\Ih \Tilde{A}}\~, \label{zpi}
\eeq
where the total first-level action $\Tilde{A}$ is
\bea
\Tilde{A}
&=&\Tilde{W}
-\frac{i\kappa\theta}{2\hbar}(\Tilde{W},\Tilde{W})\q_{\tau*(\kappa)}
+\Tilde{X}
+\frac{i\kappa\theta}{2\hbar}(\Tilde{X},\Tilde{X})\q_{\tau*(-\kappa)} \cr
&=&\Tilde{W}+\kappa\theta\Delta\q_{\tau*(\kappa)}\Tilde{W}+\Tilde{X}
-\kappa\theta\Delta\q_{\tau*(-\kappa)}\Tilde{X}
\=T\p_{(\kappa)}\Tilde{W}+T\p_{(-\kappa)}\Tilde{X}\~.
\eea
Note that the total action $\Tilde{A}$ does not contain inverse powers of
$\hbar$ due to the quantum master equations \e{tildeaddqme} for $\Tilde{W}$ 
and $\Tilde{X}$.

\section{Independence of Gauge-Fixing $\Tilde{X}$}
\label{secindependx}

\noi
The quantum BRST operator for $\Tilde{X}$ is defined as
\bea
(\sigma\p_{\Tilde{X}*}F)
&:=&\Hi e^{-\Ih \Tilde{X}}_{*}*\Delta\q_{\tau*}(e^{\Ih \Tilde{X}}_{*}*F)
-\Hi e^{-\Ih \Tilde{X}}_{*}*(\Delta\q_{\tau*} e^{\Ih \Tilde{X}}_{*})*F \cr
&=& \Hi(\Delta\q_{\tau*} F)+(\Tilde{X},F)\q_{\tau*}\~,\qquad 
F\~\in\~\cA\q_{\tau}\~,\qquad \sigma^{2}_{\Tilde{X}*}\=0\~.
\eea
Since the $\sigma\p_{\Tilde{X}*}$ operator is nilpotent, one may argue on 
general grounds that an arbitrary infinitesimal variation $\delta\Tilde{X}$ 
of the action $\Tilde{X}$ should be BRST exact,
\beq
(\sigma\p_{\Tilde{X}*}\delta\Tilde{X})\=0\~,\qquad
\delta \Tilde{X}\= (\sigma\p_{\Tilde{X}*}\delta\Psi)\~,
\eeq 
for some infinitesimal Fermion $\delta\Psi$, or equivalently,
\beq
\Ih e^{\Ih\Tilde{X}}_{*}*\delta \Tilde{X}
\=\delta e^{\Ih \Tilde{X}}_{*} 
\= \Delta\q_{\tau*}(e^{\Ih\Tilde{X}}_{*}*\delta\Psi)
-(\Delta\q_{\tau*} e^{\Ih\Tilde{X}}_{*})*\delta\Psi\~.
\eeq
By using properties \e{vartransposedop} of transposed operators, and the 
quantum master equations \e{tildeexpqme}, one may deduce that the $\Tilde{\cZ}$
partition function \e{zpi} is independent of the gauge-fixing $\Tilde{X}$. 
\beq
\delta\Tilde{\cZ}\=\int\!d\mu\~e^{\Ih\Tilde{W}}_{*(\kappa)}\cdot 
\delta e^{\Ih\Tilde{X}}_{*(-\kappa)} 
\=\int\!d\mu\~e^{\Ih\Tilde{W}}_{*(\kappa)}\cdot
\Delta\q_{\tau*(-\kappa)}
(e^{\Ih\Tilde{X}}_{*(-\kappa)}*\p_{(-\kappa)}\delta\Psi)
\=\int\!d\mu\~(\Delta\q_{\tau*(\kappa)}e^{\Ih\Tilde{W}}_{*(\kappa)})\cdot
(e^{\Ih\Tilde{X}}_{*(-\kappa)}*\p_{(-\kappa)}\delta\Psi)\=0\~.
\eeq

\section{Integrating Out The $\tau$-Extended Sector}
\label{secreducz}

\noi
One can always integrate out the new variable $\theta\!\equiv\!t^{*}_{0}$.
The boundary condition \e{bcx} creates a delta-function 
\beq
\int\! d\lambda\q_{\theta}\~e^{\Ih\cdot i\hbar\theta\lambda\q_{\theta}}
\=\int\! d\lambda\q_{\theta}\~e^{\lambda\q_{\theta}\theta}\=\delta(\theta)\~,
\eeq
and therefore one implements the condition $\theta\!=\!0$. The other new 
variable $t\q_{0}\!\equiv\!\ln(t)$ is a Schwinger proper time variable in a
world-line formalism \cite{s51}. Let us for simplicity use Darboux coordinates
$\{z^{A}_{0};t\q_{0};t^{*}_{0}\}
\!=\!\{\phi^{\alpha}_{0};\phi^{*}_{0\alpha};t\q_{0};t^{*}_{0}\}$, 
and integrate out the first-level Lagrange multipliers
$\{\lambda^{\Tilde{\alpha}}\}\!=\!\{\lambda^{\alpha};\lambda\q_{0}\}$, such
that the resulting zero-level total action $A$ is a lower truncated Laurent 
series in the $t\!\equiv\!e^{t\q_{0}}$ variable
\beq
A\=\Tilde{A}\left(\phi\q_{0};
\phi^{*}_{0}\!=\!\frac{\partial\psi}{\partial\phi\q_{0}};
t\q_{0};t^{*}_{0}\!=\!\frac{\partial\psi}{\partial t\q_{0}};
\lambda\!=\!0;\lambda\q_{0}\!=\!0;\hbar;\kappa\right)
\=\sum_{k=-M}^{\infty}A\q_{(k)}e^{k t\q_{0}}\~, \qquad 
A\q_{(k)}\=A\q_{(k)}(\phi\q_{0};\hbar;\kappa)\~.
\eeq
{}For a theory that is perturbative in the original $z$-variables, (minus) the 
lower limit is $M\!\leq\!2$. If we furthermore integrate out the Schwinger 
proper time variable $t\q_{0}$, then the $\Tilde{\cZ}$ partition function 
\e{zpi} becomes
\bea
\lefteqn{ \Tilde{\cZ}
\=\int_{-\infty}^{0}\!\!\!\!\!\!dt\q_{0}\int\![d\phi\q_{0}]\~e^{\Ih A} }\cr 
&=&\left\{\begin{array}{ll}
\frac{1}{M}
\sum_{m=0}^{\infty}\frac{1}{m!}\sum_{k\q_{1}, \ldots,k\q_{m}\geq 1-M}
\int[d\phi\q_{0}]\left(-\Ih A\q_{(-M)}\right)^{\frac{\Sigma k}{M}}
\Gamma\left(-\frac{\Sigma k}{M};-\Ih A\q_{(-M)}\right)
\prod_{i=1}^{m}\Ih A\q_{(k\q_{i})}&\for\~M\!>\!0\~,\cr 
\sum_{m=0}^{\infty}\frac{1}{m!}\sum_{k\q_{1}, \ldots,k\q_{m}\geq 1}
\frac{1}{\Sigma k}
\int[d\phi\q_{0}]e^{\Ih A\q_{(0)}}
\prod_{i=1}^{m}\Ih A\q_{(k\q_{i})}&\for\~M\!=\!0\~, \cr 
\sum_{m=0}^{\infty}\frac{1}{m!}\sum_{k\q_{1}, \ldots,k\q_{m}\geq -M}
\frac{1}{\Sigma k}
\int[d\phi\q_{0}] \prod_{i=1}^{m}\Ih A\q_{(k\q_{i})}&\for\~M\!<\!0\~,
\end{array}\right. \label{schwingerformula} \cr && 
\eea
where $\Sigma k\!:=\!\sum_{i=1}^{m}k\q_{i}$; where
$\Gamma(s;\eps)\!:=\!\int_{\eps}^{\infty}\frac{du}{u} u^{s}e^{-u}$ is the 
incomplete Gamma function; and in the case $M\!>\!0$, it has been assumed
that ${\rm Im}(A\q_{(-M)})\!>\!0$. The case $M\!<\!0$ can be viewed as the case
$M\!=\!0$ with $A\q_{(0)}\!=\!0$. The formula \e{schwingerformula} is an
expansion in Planck's constant $\hbar$ if all the subleading terms
$A\q_{(k>-M)}\!=\!\cO(\hbar)$ are quantum corrections. We stress that the
world-line path integral $\Tilde{\cZ}$ does {\em not} reproduce the standard
field-antifield path integral \cite{bv81} in the undeformed limit
$\kappa\!\to\!0$, as only the former contains a Schwinger proper time
integration.\footnote{
However, we mention an alternative procedure in the special situation where
$\kappa\Delta\q_{*}S\!=\!0$, which includes both (i) the undeformed case
$\kappa\!=\!0$ with action $\Tilde{W}\!=\!\frac{W}{t^{2}}$, and (ii) the
truncated case $\Tilde{W}\!=\!\frac{W}{t^{2}}\!=\!\lfloor S\rfloor$ with 
$\Delta\q_{*}S\!=\!0$. In these two cases, shift the $\Tilde{W}$ action with a
one-loop contribution
$\Tilde{W}=\frac{W}{t^{2}}\longrightarrow 
\Tilde{W}=\frac{W}{t^{2}}+i\hbar\ln(1\!-\!t^{2})
=\frac{W}{t^{2}}+\Hi\sum_{k=1}^{\infty}\frac{t^{2k}}{k}$.
One may check that the shifted $\Tilde{W}$ action also satisfies the
quantum master equation \e{tildeaddqme}. Now choose the $t$ integration
contour as a small circle around $t\!=\!1$. The one-loop correction 
$\oint_{1}\!\frac{dt}{t}\~e^{\Ih\cdot i\hbar\ln(1-t^{2})}
=-\oint_{1}\!\frac{dt}{t}\frac{1}{t\!+\!1}\frac{1}{t\!-\!1}$ creates a simple
pole at $t\!=\!1$, and thereby one implements the condition $t\!=\!1$.
Therefore the $\Tilde{\cZ}$ path integral \e{zpi} reduces (up to a constant
multiplicative factor) to the standard $W$-$X$-form
$\Tilde{\cZ}=\int[dz][d\lambda]\~e^{\Ih(W+X)}=\cZ$. In the undeformed case
$\kappa\!=\!0$, the $W$ action \e{wsm01} at $t\!=\!1$ becomes the standard loop
expansion, which satisfies the standard quantum master equation
$\Delta e^{\Ih W}\!=\!0$.}

\vspace{0.8cm}

\noi
{\sc Acknowledgement:}~
I.A.B.\ would like to thank M.~Lenc, R.~von Unge and the Masaryk University 
for the warm hospitality extended to him in Brno. K.B.\ would like to thank
M.~Vasiliev, the Lebedev Physics Institute and the Erwin Schr\"odinger
Institute for warm hospitality. The work of I.A.B.\ is supported by grants
RFBR 08--01--00737, RFBR 08--02--01118 and LSS--1615.2008.2. The work of K.B.\
is supported by the Ministry of Education of the Czech Republic under the
project MSM 0021622409. 

\appendix

\section{Proof of \eq{deformedtaudelta}}
\label{appdeformedtaudelta}

\bea
T^{-1}\Delta\q_{\tau}T&=&\Delta\q_{\tau}T
\=\Delta\q_{\tau}(1+\kappa\theta\Delta\q_{\tau*})
\=\Delta\q_{\tau}+\kappa[\Delta\q_{\tau},\theta]\Delta\q_{\tau*}
\=\Delta\q_{\tau}+K\q_{\tau}\Delta\q_{\tau}\frac{1}{1\!-\!K\q_{\tau}} \cr
&=&\Delta\q_{\tau}\frac{1}{1\!-\!K\q_{\tau}}\=\Delta\q_{\tau*}\~.
\label{deformedtaudelta02} 
\eea

\section{Proof of \eq{ttrelation}}
\label{appttrelation}

\bea
T^{-1}T&:=&(1-\kappa\theta\Delta\q_{\tau})(1+\kappa\theta\Delta\q_{\tau*})
\=1-\kappa\theta\Delta\q_{\tau}+\kappa\theta\Delta\q_{\tau*}
-\kappa^{2}\theta[\Delta\q_{\tau},\theta]\Delta\q_{\tau*} \cr
&=&1-\kappa\theta\Delta\q_{\tau}
+\kappa\theta\Delta\q_{\tau}\frac{1}{1\!-\!K\q_{\tau}}
-\kappa\theta K\q_{\tau}\Delta\q_{\tau}\frac{1}{1\!-\!K\q_{\tau}}
\=1\~.\label{ttrelation02} \\
TT^{-1}&:=&(1+\kappa\theta\Delta\q_{\tau*})(1-\kappa\theta\Delta\q_{\tau})
\=1-\kappa\theta\Delta\q_{\tau}+\kappa\theta\Delta\q_{\tau*}
-\kappa^{2}\theta[\Delta\q_{\tau}\frac{1}{1\!-\!K\q_{\tau}},\theta]
\Delta\q_{\tau} \cr
&=&1-\kappa\theta\Delta\q_{\tau}
+\kappa\theta\Delta\q_{\tau}\frac{1}{1\!-\!K\q_{\tau}}
-\kappa\theta K\q_{\tau}\frac{1}{1\!-\!K\q_{\tau}}\Delta\q_{\tau}
\=1\~.\label{ttrelation03}
\eea

\section{Proof of \eq{deformtauantibracket}}
\label{appdeformtauantibracket}

\bea
(B,B)\q_{\tau*}&:=&T^{-1}(TB,TB)\q_{\tau}
\=(1-\kappa\theta\Delta\q_{\tau})(TB,TB)\q_{\tau}\=I-II \cr
&=&(B,B)\q_{\tau}+2(\Delta\q_{\tau*}B)\cdot(K\q_{\tau}B)  
\~,\qquad B\~\in\~\cA\q_{\tau}\~,\qquad \eps(B)\=0\~,
\label{deformtauantibracket02}
\eea
where
\bea
I&:=&(TB,TB)\q_{\tau}
\=(B+\kappa\theta\Delta\q_{\tau*}B,B
+\kappa\theta\Delta\q_{\tau*}B)\q_{\tau} \cr
&=&(B,B)\q_{\tau}-2\kappa(\Delta\q_{\tau*}B)\cdot(\theta,B)\q_{\tau}
+2\kappa\theta(\Delta\q_{\tau*}B,B)\q_{\tau}
+2\kappa^{2}\theta(\Delta\q_{\tau*}B,\theta)\q_{\tau}
\cdot\Delta\q_{\tau*}B\~, \\
II&:=&\kappa\theta\Delta\q_{\tau}(TB,TB)\q_{\tau}
\=2\kappa\theta(\Delta\q_{\tau}TB,TB)\q_{\tau}
\=2\kappa\theta(\Delta\q_{\tau}B
+\kappa[\Delta\q_{\tau},\theta]\Delta\q_{\tau*}B,\~TB)\q_{\tau} \cr
&=&2\kappa\theta(\Delta\q_{\tau}B
+K\q_{\tau}\Delta\q_{\tau}\frac{1}{1\!-\!K\q_{\tau}}B,\~TB)\q_{\tau}
\=2\kappa\theta(\Delta\q_{\tau*}B,\~B+\kappa\theta\Delta\q_{\tau*}B)\q_{\tau}
\cr 
&=&2\kappa\theta(\Delta\q_{\tau*}B,B)\q_{\tau}
+2\kappa^{2}\theta(\Delta\q_{\tau*}B,\theta)\q_{\tau}\cdot\Delta\q_{\tau*}B\~.
\eea
Now use polarization of \eq{deformtauantibracket02} to prove
\eq{deformtauantibracket}, \cf \eg \Ref{b06}.

\section{Proof of \eq{trivstarproduct}}
\label{apptrivstarproduct}

\bea
B*B&=&T^{-1}(TB)^{2}\=T^{-1}(B+\kappa\theta(\Delta\q_{\tau*}B))^{2}
\=(1-\kappa\theta\Delta\q_{\tau})
\left(B^{2}+2\kappa\theta B\Delta\q_{\tau*}B\right) \cr
&=&I-II-III\=B^{2}-\kappa\theta(B,B)\q_{\tau}
-2\kappa\theta(K\q_{\tau}B)\cdot\Delta\q_{\tau*}B \cr
&=&B^{2}-\kappa\theta(B,B)\q_{\tau*}\~,\qquad
B\~\in\~\cA\q_{\tau}\~,\qquad \eps(B)\=0\~, \label{trivstarproduct02}
\eea
where
\bea
I&:=&B^{2}+2\kappa\theta B\Delta\q_{\tau*}B
\=B^{2}+2\kappa\theta B\Delta\q_{\tau}\frac{1}{1\!-\!K\q_{\tau}}B\~, \\
II&:=&\kappa\theta\Delta\q_{\tau}(B^{2})
\=2\kappa\theta B\Delta\q_{\tau}B+\kappa\theta(B,B)\q_{\tau}\~, \\
III&:=&2\kappa^{2}\theta\Delta\q_{\tau}\theta B\Delta\q_{\tau*}B
\=2\kappa^{2}\theta[\Delta\q_{\tau},\theta]B\Delta\q_{\tau*}B
\=2\kappa\theta K\q_{\tau}B\Delta\q_{\tau*}B \cr
&=&2\kappa\theta (K\q_{\tau}B)\cdot\Delta\q_{\tau*}B
+2\kappa\theta BK\q_{\tau}\Delta\q_{\tau}\frac{1}{1\!-\!K\q_{\tau}}B\~.
\eea
Now use polarization of \eq{trivstarproduct02} to prove \eq{trivstarproduct}.

\section{Proof of \eq{starexpdef}}
\label{appstarexpdef}

\bea
e^{B}_{*}&=&T^{-1}e^{(TB)}\=T^{-1}e^{B+\kappa\theta(\Delta\q_{\tau*}B)}
\=(1-\kappa\theta\Delta\q_{\tau})e^{B}(1+\kappa\theta\Delta\q_{\tau*}B) \cr
&=&I-II-III\=e^{B}\left(1-\Hf\kappa\theta(B,B)\q_{\tau}
-\kappa\theta(K\q_{\tau}B)\cdot\Delta\q_{\tau*}B\right)\cr
&=&e^{B}\left(1-\Hf \kappa\theta(B,B)\q_{\tau*}\right)
\~,\qquad B\~\in\~\cA\q_{\tau}\~,\qquad \eps(B)\=0\~,\label{starexpdef02}
\eea
where
\bea
I&:=&e^{B}(1+\kappa\theta\Delta\q_{\tau*}B)
\=e^{B}(1+\kappa\theta\Delta\q_{\tau}\frac{1}{1\!-\!K\q_{\tau}}B)\~, \\
II&:=&\kappa\theta(\Delta\q_{\tau}e^{B})
\=\kappa\theta e^{B}\left(\Delta\q_{\tau}B+\Hf(B,B)\q_{\tau}\right)\~, \\
III&:=&\kappa^{2}\theta\Delta\q_{\tau}\theta e^{B}\Delta\q_{\tau*}B
\=\kappa^{2}\theta[\Delta\q_{\tau},\theta]e^{B}\Delta\q_{\tau*}B
\=\kappa\theta K\q_{\tau}e^{B}\Delta\q_{\tau*}B \cr
&=&\kappa\theta e^{B}(K\q_{\tau}B)\cdot\Delta\q_{\tau*}B
+\kappa\theta e^{B}K\q_{\tau}\Delta\q_{\tau}\frac{1}{1\!-\!K\q_{\tau}}B\~.
\eea

\end{document}